\documentclass{article}

\usepackage{arxiv}

\usepackage[utf8]{inputenc} 
\usepackage[T1]{fontenc}    
\usepackage{hyperref}       
\usepackage{url}            
\usepackage{booktabs}       
\usepackage{amsfonts}       
\usepackage{nicefrac}       
\usepackage{microtype}      
\usepackage{lipsum}		
\usepackage{graphicx}
\usepackage{doi}

\newtheorem{definition}{Definition}

\usepackage[
]{biblatex}
\title{Attributed Stream Hypergraphs: temporal modeling of node-attributed high-order interactions}

\author{ Andrea Failla \\
	Department of Computer Science\\
	University of Pisa\\
	\texttt{andrea.failla@phd.unipi.it} \\
	\And
	Salvatore Citraro \\
        KDD-LAB\\
	ISTI-CNR\\
	\texttt{salvatore.citraro@isti.cnr.it} \\
        \And
        Giulio Rossetti \\
        KDD-LAB\\
        ISTI-CNR\\
        \texttt{giulio.rossetti@isti.cnr.it}
}

\renewcommand{\shorttitle}{Attributed Stream Hypergraphs}

\begin{document}
\maketitle

\begin{abstract}
Recent advances in network science have resulted in two distinct research directions aimed at augmenting and enhancing representations for complex networks.
The first direction, that of high-order modeling, aims to focus on connectivity between sets of nodes rather than pairs, whereas the second one, that of feature-rich augmentation, incorporates into a network all those elements that are driven by information which is external to the structure, like node properties or the flow of time.
This paper proposes a novel toolbox, that of Attributed Stream Hypergraphs (ASHs), unifying both high-order and feature-rich elements for representing, mining, and analyzing complex networks.
Applied to social network analysis, ASHs can characterize complex social phenomena along topological, dynamic and attributive elements.
Experiments on real-world face-to-face and online social media interactions highlight that ASHs can easily allow for the analyses, among others, of high-order groups' homophily, nodes' homophily with respect to the hyperedges in which nodes participate, and time-respecting paths between hyperedges. 
\end{abstract}


\section{Introduction}

Complex networks provide a lens through which to illustrate plenty of behaviors that characterize humans as social animals.
The elements of graph theory constituted the most helpful toolbox to represent and analyze social networks, with the intention to study complex behavior by mapping any possible kind of human contact, interaction, or relation as pairs of edges between unit elements called nodes.
Network science, founded on such a basis, has been able to unravel many social patterns hidden at several scales of human relationships.
Global network structures such as rich-clubs \cite{colizza2006detecting} and core-periphery structures \cite{gallagher2021clarified}, together with meso-scale organizations in blocks or communities \cite{fortunato2016community}, give an idea to the extent to which graphs are useful to grasp the knowledge of complex social architectures.
However, the intrinsic nature of graphs to map dyadic patterns does not allow encoding explicitly group connectivity or high-order relations, which are fundamental in the social sphere.
An increasing number of works recently started to address the mathematical tools of hypergraph theory \cite{aksoy2020hypernetwork} and simplicial complexes \cite{iacopini2019simplicial, battiston2020networks} to implement multi-body representations of social systems \cite{torres2021and}.
Such new lines of hyper-network science aim to point out the importance of high-order interactions when studying the social dynamics of groups \cite{veldt2023combinatorial, sarker2023generalizing} or nodes embedded in groups rather than within neighborhoods built upon pairwise connections \cite{failla2023attributed}.

Parallel to this new interest in augmented topologies, other lines of research in network science focus on representations combining the structure with the large amount of domain-specific elements often available from a social system, like people's qualities or preferences, or with any kind of information external to the system that can be related to the structure, e.g., the flow of time that could affect topological changes. 
Mining such semantically augmented networks helps to unhearth many interesting social properties, from assortative mixing patterns based on common preferences \cite{newman2003mixing} to the rules hidden in the formation and evolution of groups \cite{palla2007quantifying, rossetti2018community}.
The term `\textit{feature-rich}' networks \cite{interdonato2019feature} unifies all these augmented implementations that aim to add external, semantic information to a complex structure.
Originally designed for pairwise networks, we believe that any complex topology could benefit from a feature-rich implementation, thus also networked representations built upon hypergraphs and simplicial complexes.

Hence, the objective of this work is to address the analysis of high-order patterns together with feature-rich elements.
Generalizing the feature-rich framework, we aim to represent and analyze complex social phenomena along the following three dimensions: topology, dynamic features, and node attributes.
To this purpose we introduce \textit{ASH}, an \textit{A}ttributed \textit{S}tream-\textit{H}ypernetwork implementation for representing high-order temporal networks with attributive information on nodes.

The rest of the work is organized as follows.
Section \ref{sec:related} sums up the principal literature on the three main complex network contexts surrounding this work, namely the dynamic, the node-attributed, and the high-order representations for networks.
Section \ref{sec:ash} introduces a formalism for the Attributed Stream-Hypergraph, our framework for addressing node-attributed evolving high-order topologies.
Section \ref{sec:exp} discusses our main results on real-world scenarios, from face-to-face contacts to user interactions on online platforms.
Section \ref{sec:disc} concludes the work.
Finally, in the Appendix we introduce a Python library to work with Attributed Stream Hypergraphs.

\section{Related Work}
\label{sec:related}
In the following, we provide an overview of the main enriched/augmented network implementations that are addressed in the work.
First, we discuss dynamic and node-attributed network representations; then, we sum up the emerging contributions about high-order representations for complex systems. 
\\ \ \\
\noindent \textbf{Dynamics of networks}.
Many network data that represent human activity have an intrinsic dynamic nature, from e-mail exchanges \cite{klimt2004introducing} and financial transactions \cite{zhao2018stock}, which are instantaneous forms of connections, to face-to-face interactions, that involve a certain duration, and friendships, that are generally stable and persistent over time.
Hence, choosing a proper representation for modeling the dynamics of all these different social behaviors is not a straightforward task.
Different temporal semantics impose different representations \cite{holme2012temporal,rossetti2018community}, being possible to categorize them according to the following properties: i) stability, e.g., when dynamic data are represented as a snapshot sequence from a time-window aggregation \cite{ribeiro2013quantifying,chiappori2021quantitative}; ii) duration, e.g., when data are represented as interval graphs \cite{holme2012temporal}; and iii) immediacy, e.g., when data are represented as a stream graph of temporal nodes and connections \cite{latapy2018stream}.
In this work, we will mainly focus on such stream graphs, that have been proven to extend and generalize classic centrality measures \cite{simard2021computing}, and multi-layer structure as well \cite{parmentier2019introducing}.
More generally, among the most interesting and cutting-edge analyses on dynamic networks, we can mention community detection \cite{rossetti2018community}, link prediction \cite{divakaran2020temporal}, and mixing pattern estimation \cite{citraro2022delta}, as well as works extending properties like reciprocity \cite{chowdhary2023temporal} and structures like rich-clubs \cite{pedreschi2022temporal} to dynamic environments.
\\ \ \\
\noindent \textbf{Networks with attributes}.
Attributes or metadata often describe the properties of the nodes involved in networked data.
Node attributes can be fruitfully used for improving results on classic network tasks, e.g., in community detection, where both tight connectivity and label homogeneity within communities need to be guaranteed \cite{chunaev2020community}.
Attribute-enriched implementations can support analyses on the combined structural and attributive dimensions, searching for possible relations between the properties of nodes and how they are likely to connect \cite{mcpherson2001birds, newman2003mixing}. 
Node attributes can be leveraged for estimating homophily and heterogeneous mixing patterns \cite{peel2018multiscale,rossetti2021conformity}.
Other tasks oriented to machine learning points of view can leverage on node metadata -- e.g., the distribution of values within the adjacent neighborhood of a target node -- for node classification and link prediction purposes \cite{bhagat2011node}.
There is also an emerging effort toward the exploration of global patterns of connectivity in attributed data, which still is an unexplored topic in the literature on feature-rich networks.
Attributed backboning, for instance, is the task of finding the subtree of a graph that spans over the nodes with a minimized connection cost, where such cost is determined by node affinitive attributes \cite{guan2019attribute}.
Similarly, a (k,r)-core structure is a subgraph that is cohesive with respect to both node connectivity and similarity \cite{zhang2017when}.
\\ \ \\
\noindent \textbf{High-order networks}.
Although traditional network science mostly addressed pairwise network representations, many dynamics can be better thought of as high-order representations involving relations between groups of nodes.
As an emerging line of research \cite{battiston2020networks,joslyn2020hypernetwork,torres2021and}, the expressive power of such high-order relations is yet largely unexplored.
The interest in the physics of high-order interactions is growing \cite{battiston2021physics}, being extensively explored in the area of diffusive processes on networks, e.g., for studying social contagion with simplicial complexes \cite{iacopini2019simplicial}, in time-varying settings as well \cite{chowdhary2021simplicial}.
High-order structures varying in time are an important and emerging trend of research \cite{cencetti2021temporal, comrie2021hypergraph}.
They have been applied to study the network structure of scientific revolutions \cite{ju2020network}, or the evolution of high-order linguistic networks in scientific texts \cite{christianson2020architecture}.
There is also an increasing interest in the analysis of high-order interactions with attributes, e.g., measures for estimating homophily in hypergraphs and simplicial complexes \cite{veldt2023combinatorial, sarker2023generalizing}, or integrating node attributes through annotated high-order models \cite{chodrow2020annotated}.
The high-order structure of static/dynamic networks is often addressed by investigating datasets originally designed for graph-based analysis, thus one of the most intriguing future challenges is the inference of statistically significant high-order interactions from complex systems \cite{musciotto2021detecting}.
Finally, some lines of works tend to be more conservative, as in the case of the s-line graph analysis for hypergraphs \cite{aksoy2020hypernetwork}, where the hyperedge-projection of the hypergraph is used to apply, for instance, graph-based centrality measures to characterize hyperedges rather than nodes.

\section{Attributed Stream Hypergraphs}
\label{sec:ash}

To study dynamic high-order social interactions, simply borrowing results from the existing literature is not enough. 
Hypergraphs and/or simplicial complex has been not adequately defined in the presence of evolving topologies.
Moreover, individuals embedded in a social system can often be characterized by multiple features --- \emph{profiles} that contextualize some of the key properties playing a role in social interactions (e.g., nationality, gender, age\dots).
In this Section we introduce ASH, our Attributed Stream Hypergraph model, adequately defined for evolving high-order interactions with semantically enriched nodes.
We formally define ASHs as follows:

 \begin{definition}[ASH]
 Let $\mathcal{S}=(T,V,W,E,L)$ be a stream hypergraph, where:
\begin{itemize}
    \item $T = [\mathrm{A}, \Omega]$ is the set of discrete time instants, with $\mathrm{A}$ and $\Omega$ the initial and final instants;
    \item $V$ is the set of the nodes of the temporally flattened hypergraph, namely the set of all nodes appearing during the ASH's lifespan;
    \item $W \subseteq T \times V$ is the set of temporal nodes;
    \item $E \subseteq T \times V^n$ is the set of temporal hyperedges such that $(t,N) \in E$ implies that $N \subseteq V$ and $\forall \: u_i \in N,  (t,u_i) \in W$;
    \item $L=\{l_1,...l_m\}$ is the set of $m$ node attributes such that $l_{(t,u)}$ with $(t,u) \in W$ and $t \in T$, identifies the categorical value of the attribute $l$ associated to $u$ at time $t$.
\end{itemize}
\end{definition}

\noindent ASHs bring together high-order interactions, temporal dynamics, and node attributes.
It should be noted that other modeling frameworks can be thought of as particular instances of an ASH, where one of the three dimensions is \textit{switched off}.
For instance, given an ASH $\mathcal{S}=(T,V,W,E,L)$, it is possible to switch off a dimension that results in one of the following representations:
\begin{itemize}
    \item an attributed stream graph \cite{citraro2022delta} for $|N| = 2, \forall \: (t, N) \in E$, where $|N|$ identifies the number of nodes included in hyperedge $(t, N) \in E$;
    \item a static node-attributed hypergraph \cite{veldt2023combinatorial} for $|T| = 1$ (i.e., there is no temporal dynamics), which implies $W = V$ and $E \subseteq V^n$;
    \item a stream hypergraph (without node attributes) for $L = \emptyset$.
\end{itemize}

\subsection{Inheriting from stream graphs and hypergraphs}

ASHs are a conservative extension of stream graphs \cite{latapy2018stream} and hypergraphs \cite{battiston2020networks, aksoy2020hypernetwork}, thus inheriting from such frameworks their peculiar concepts.
For instance, ASHs inherit from stream graphs the peculiarities of temporal nodes and temporal edges, since the nature of nodes and edges is analyzed with respect to the times they appear in the temporal stream.
Nodes/edges can be thought of as temporal entities that can be present or absent at a certain time in the stream, so that the \textit{contribution} of a node/edge is said to be \textit{equal to 1} -- i.e., represented as a whole quantity -- only if it is present all the time in the stream.
With a rapid example, the contribution of an edge $uv$ is computed as follows: $m_{uv} = \frac{|T_{uv}|}{|T|}$, where $|T_{uv}|$ represents the number of time instants where $uv$ is present, and $|T|$ is the overall number of time instants.
Naturally, the main difference with stream graphs is that, in an ASH, the temporal presence of an interaction is accounted for hyperedges. 
This aspect captures the fact that nodes/edges might not be present all the time, thus $|W|$, the sum of active nodes across all temporal instants, and $|T \times V|$, the sum of all possible active nodes across all temporal instants, might differ significantly.
The contribution of temporal hyperedges is computed under the same rationale, i.e., the sum of active hyperedges across all temporal instants over the sum of all possible active hyperedges across all temporal instants.
Finally, in the case when all nodes are present all the time in the stream, the representation is called \textit{link stream}, and it is a possibility allowed for ASHs as well.
\\ \ \\
Another key concept that can be generalized to ASHs is that of path.
Paths on graphs have already been extended to hypergraphs within the \textit{s}-analysis framework \cite{aksoy2020hypernetwork}.
This frameowrk builds on the idea that hyperedge paths (or any walk, equivalently) not only have a \textit{length}, i.e., the number of hyperedges crossed during the walk, but also a \textit{width}, i.e., the cardinality of the minimum intersection between subsequent hyperedges. 
For instance, an \textit{s}-walk of width 3 (\textit{3-walk}, equivalently) is a sequence of hyperedges where each edge intersects on at least 3 nodes with its predecessor (except for the hyperedge at the beginning) as well as with its successor (except for the hyperedge at the end).
However, the dynamic nature of ASHs comes with the added constraint of temporal contiguity.
In other words, in a temporal setting, each subsequent hyperedge along an \textit{s}-walk must come with non-decreasing, adjacent time instants. 
This also implies that, aside from length and width, a temporal \textit{s}-walk also has a \textit{duration}, namely the number of time instants occurring between the beginning and the end of the walk.
Hence, we define a \textit{time-respecting s-walk} as follows:

\begin{definition}[Time-respecting s-walk]
    A time-respecting \textit{s}-walk of length \textit{k}, width \textit{s}, and duration \textit{d} is a sequence $P = \{(t_0, N_0), (t_1, N_1), \dots, (t_{k-1}, N_{k-1})\}$ such that:
    \begin{itemize}
        \item $P \subseteq E$
        \item $t_i \leq t_{i+1}$ for all \textit{i}s, where $i \in \mathbf{Z^+} \land i < k$ identifies the position of a hyperedge along the walk;
        \item $s \in \mathbf{Z^+} \land s \leq |(N_i) \cap (N_{i+1})|$;
        \item $d = t_{k-1} - t_0$;
    \end{itemize}
\end{definition}

\noindent By leveraging the above formulation, the notions of shortest, fastest, fastest-shortest, shortest-fastest, and foremost time-respecting \textit{s}-walks can be deduced as already done for stream graphs \cite{latapy2018stream}, e.g., shortest paths are the ones with minimal length $k$, fastest paths are the ones with minimal duration $t_k - t_0$, fastest-shortest are the fastest paths
among the shortest ones, and shortest-fastest, viceversa; foremost paths, independently from length and duration, are the ones that reach first the destination.
\\ \ \\
Another concept that can be extended dynamically is that of node's \textit{star}, namely the set of hyperedges where the node is present.
This can be limited to include only hyperedges that are active at a specific point in time.
\begin{definition}[Temporal Star]
    Let $u \in V$ be a node in the ASH. 
    The temporal star of $u$ at time $t$ is the set of temporal hyperedges that include $u$ in $t$, and is denoted $D(t,u) = \{(t, N): (t, N) \in E \land u \in N\}$.
\end{definition}

\noindent Temporal star analysis allows quantifying node-level properties of temporal high-order structures \cite{comrie2021hypergraph}, as well as eventually extending to the temporal dimension concepts like hyperego-network density and overlap \cite{lee2021hyperedges}.

\subsection{Towards temporal mixing patterns estimation}

Apart from combining the stream graph's evolutionary nature with the hypergraph's high-order structure, ASHs can integrate time-evolving node attributes, i.e., labels that (may) change in time.
This peculiarity allows studying not only how individuals' characteristics change (e.g., opinions, political leaning) but also how such changes relate/affect the topological structure surrounding them.
As a node's attribute values might vary through time, one can quantify the extent of their \textit{consistency}, namely to what extent a node's attribute value remains constant over time.

\begin{definition}[Consistency]
    Let $u \in V$ be a node, and $l \in L$ be an attribute such that $l_{(t, u)}$ denotes the attribute value of $u$ at time $t$; 
    let $T_u$ identify the set of time instants where $u$ is present. 
    The $Consistency$ of $u$ with regards to $l$ ranges in $[0,1]$ and is computed as:
    \begin{equation}
    \label{eq:consistency}
        Consistency(u, l) = 1 - (-\sum_{t \in T_u} p(l_{(t,u)}) \log p(l_{(t,u)})).
    \end{equation}
\end{definition}

\noindent Consistency can be extended to the whole ASH by computing the average as follows:
\begin{equation}
\label{eq:avgconsistency}
    Consistency(\mathcal{S}, l) =\frac{1}{|V|} \sum_{u \in V} Consistency(u, l),
\end{equation}
which quantifies the extent to which an ASH's nodes keep their attribute value constant in time. 
\\ \ \\
\noindent Henceforth, we may be interested in quantifying (a) hyperedges' homogeneity, which is a rising hot topic in attributed high-order analyses \cite{veldt2023combinatorial, sarker2023generalizing}, and also (b) the level of homogeneity of a target node with respect to the set of hyperedges it belongs.
In case (a), several options are possible, e.g., as cleanly proposed in \cite{veldt2023combinatorial, sarker2023generalizing} with statistically validated measures.

More straightforward ways to measure hyperedges' homogeneity can consist in finding an aggregate value that sums up the \textit{characteristics} of a hyperedge with respect to the labels carried by the nodes within it. 
For instance, a characteristic value can be the frequency of the most frequent class within a hyperedge.
Hence, we can use hyperedges' purity \cite{citraro2020identifying} as follows:

\begin{definition}[Temporal Purity]
    Let $(t, N) \in E$ be a temporal hyperedge and $l \in L$ be a node attribute.
    Let $max_{l\in L} (\sum_{n \in N} l_{(t,n)})$ be the most frequent categorical value within $(t,N)$. 
    The temporal purity of $(t,N)$ is the relative frequency of the most frequent value and ranges in $[0,1]$:
    \begin{equation}
    \label{eq:purity}
        Purity(t,N,l) = \frac{\max_{l\in L} (\sum_{n \in N} l_{(t,n)})}{|(t,N)|},
    \end{equation}
\end{definition}

\noindent Similarly, another characteristic value that can be used to describe a hyperedge is entropy, which quantifies the degree of disorder related to the nodes' attribute values within the hyperedge. 

\begin{definition}[Entropy]
    Let $(t, N) \in E$ be a temporal hyperedge and $l \in L$ be a node attribute.
    Let $A_{(t, N), l}$ be the set of the attribute values of $l$ in $(t,N)$. 
    The entropy of $(t,N)$ with respect to $l$ ranges in $[0,1]$ and is computed as follows:
    \begin{equation}
    \label{eq:entropy}
        H(t, N, l) = - \sum_{i}^{|A_{(t,N),l}|} p(i) \log p(i)
    \end{equation}
\end{definition}

\noindent In case (b), our focus is on a target node $u \in V$ aiming to analyze $u$'s homogeneity with respect to its attribute value $l_{(t,u)}$.
We can still associate each hyperedge in $D(t,u)$ with a characteristic value.
The ones described so far, i.e., purity and entropy, result in continuous values.
However, we can also characterize a hyperedge by means of a categorical value.
Here, for instance, we describe each hyperedge in the temporal star of a target node $u$ by means of the most frequent attribute value within the hyperedge, namely $max_{l\in L} (\sum_{n \in N} l(t,n))$, with $(t,N) \in D(t,u)$. 
Having such categorical value can allow us to compute the relative frequency of such characteristic values with respect to the label of the target node.

\begin{definition}[Homogeneity]
    Let $u \in V$ be a node with attribute value $l_{(t, u)}$, $l \in L$.
    Let be $D(t, u)$ the temporal star of $u$, and $max_{l\in L} (\sum_{n \in N} l_{(t,n)})$, with $(t,N) \in D(t,u)$ the most frequent categorical value of a hyperedge belonging to the star of $u$.
    The star homogeneity of $u$ with respect to $l_{t,u}$ is the relative frequency of the hyperedges in $D(t,u)$ that share with $u$ its same attribute value $l_{t,u}$. It ranges in $[0,1]$ and is defined as follows:
    \begin{equation}
    \label{eq:sthom}
        Homogeneity(t, u, l) =  \frac{
        |\{(t,N) \in D(t, u) : max_{l\in L} (\sum_{n \in N} l_{(t,n)}) = l(t,u)\}|
        }{|D(t, u)|},
    \end{equation}
    \label{eq:homogeneity}
\end{definition}

\noindent Star homogeneity quantifies a node's degree of embeddedness across all of the contexts/interactions it finds itself in.

Finally, purity, entropy, and star homogeneity, can be averaged to capture the global behavior of the ASH:

\begin{equation}
\label{eq:avgpurity}
    Purity(\mathcal{S}, l) = \frac{1}{|E|} \sum_{(t,N) \in E}Purity(t, N, l).
\end{equation}

\begin{equation}
\label{eq:avgentropy}
    H(\mathcal{S}, l) = \frac{1}{|E|} \sum_{(t,N) \in E}H(t, N, l)
\end{equation}

\begin{equation}
\label{eq:avgsthom}
    Homogeneity(\mathcal{S},l) = \frac{1}{|V_t|} \sum_{u \in V}Homogeneity(t, u, l).
\end{equation}

\noindent Note that Eq. \ref{eq:avgsthom} averages over the number of nodes, as star homogeneity quantifies node behavior; conversely, Eqq. \ref{eq:avgpurity} and \ref{eq:avgentropy} average over the number of hyperedges.

\section{Experiments}
\label{sec:exp}

In this Section we provide basic experiments to test ASH's potentialities.
We define the two following case studies:

\begin{itemize}
    \item We provide a characterization of face-to-face interactions within the SocioPatterns project\footnote{\url{www.sociopatterns.org}}, focusing particularly on children in a primary school \cite{stehle2011high}, and medical staff and patients in a hospital ward \cite{vanhems2013estimating};
    \item We build a case study on online social network discussions about political topics \cite{morini2021toward} to describe aspects related to pairwise-based vs. high-order-based representations.
\end{itemize}

We promote two different analyses coherently with the different nature/persistence of connectivity in face-to-face contacts and online discussions.
In face-to-face interactions we are more interested in analyzing temporal mixing patterns and time-respecting paths with different aggregation windows.
In online discussions, where more stability can be reached in users' intreraction, we promote a comparison between pairwise and high-order representations in characterizing users' mixing patterns.
The experiments are conducted by leveraging our Python library handling ASH-structured data\footnote{\url{https://github.com/GiulioRossetti/ASH}}, which is introduced and discussed in the Appendix.

\subsection{High-order Temporal Dynamics in Face-to-Face Contacts}
As mentioned, we analyze the dynamics of children in a primary school and individuals in a hospital ward.
Some detailed information is provided as follows:

\begin{itemize}
    \item \textit{Primary School} \cite{stehle2011high}: this dataset contains face-to-face interactions between children during the whole school day: node metadata include children's gender and class;
    \item \textit{Hospital} \cite{vanhems2013estimating}: this dataset contains the temporal contact data between medical doctors (MED), nurses and paramedics (NUR), administrative staff (ADM), and patients (PAT) in a short-stay geriatric unit of a University hospital. Data were collected for a week.
\end{itemize}

For representing the temporal higher-order structure, we leverage a similar method as that introduced in \cite{cencetti2021temporal}, namely: if at time $t$ there are $n*(n+1)/2$ dyads between the members of a set of $n$ nodes such that they are involved in a fully connected clique, such links are promoted to form a $n$-hyperedge.
We aim to characterize the hypernetworks with regard to their structure, node features, and dynamics, at different time aggregations (1 minute, 5 minutes, 10 minutes, 30 minutes, and 1 hour).
\begin{figure}[]
\includegraphics[width=0.9\textwidth]{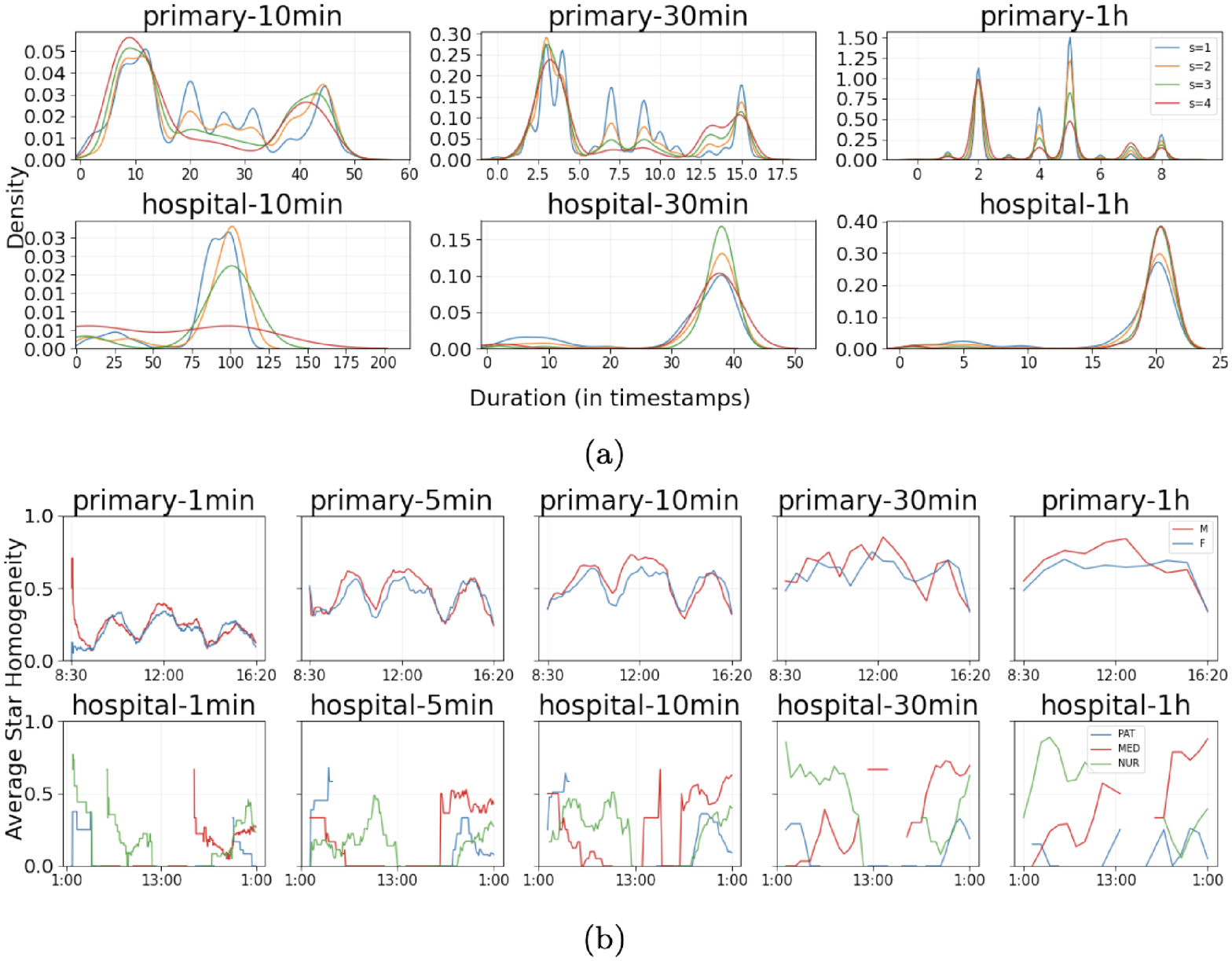}
    \caption{Duration of shortest s-paths (a) and temporal trends of average star homogeneity (b) for Primary School and Hospital Ward at different aggregation windows.}
  \label{fig:sp}
\end{figure}

\subsubsection{Time-respecting paths in face-to-face contacts}

ASHs allow to define time-respecting paths between incident hyperedges.
As mentioned, hypergraph paths add the notion of width, thus we could be interested in observing how much this parameter affects the length/duration of the walks.
Figure \ref{fig:egos} (a) shows the distribution of the duration of shortest s-paths (for s equal to 1, 2, 3, and 4) for both primary school and hospital ward at different aggregation windows.
As expected, for s=1, we observe a larger number of walks, and this difference is more visible when the aggregation window is smaller (10 minutes).
In primary school several peaks are visible, highlighting the fact that several events are present during the whole school day, like lessons in classes and several breaks.
Interestingly, larger aggregation windows in the hospital ward show that the majority of hyperedges tend to be reached distant in time, intuitively related to the fact that patients can be reached only through the interaction between nurse and medical staff.

\subsubsection{Temporal mixing patterns in face-to-face contacts}

ASHs allow to study high-order dynamic mixing patterns.
In the Sociopatterns datasets we can estimate them at different temporal scales.
Figure \ref{fig:egos} (b) describes the temporal trends of average star homogeneity (cf. Eq. \ref{eq:homogeneity}) for the two hypernetworks at different aggregation windows.
Clear temporal patterns emerge in the primary school dataset, where the attribute observed is children's gender.
Children's interactions are more randomly mixed during lesson breaks, and no differences emerge between male and female behavior.
In the hospital ward, nurses' interactions are homophilic at late-night/early morning, while MDs' are more homophilic in the evening.
The temporal aggregation windows have an impact on both datasets.
In primary school, larger windows let more homogeneous interactions emerge.
However, larger windows do not let us distinguish between class hours and breaks.
Conversely, larger windows in the hospital ward let us observe that some categories are more homophilic than others, e.g., nurses, while patients tend to be disassortative all the time, coherently with the fact that they stay in different rooms \cite{vanhems2013estimating} and they are visited only by nurses and medical staff.

\subsection{Homophilic Behaviors in Pairwise and Group Political Discussions on Reddit}

We focus on data collected from the debate between Trump supporters and anti-Trump citizens during the first two and half years of Donald Trump’s presidency, covering a period between January 2017 and July 2019.
The debates cover both controversial/polarizing sociopolitical issues and broader discussions within the US political ideologies, as follows:

  \begin{itemize}
        \item \textit{Gun Control}: this topic is identified by collecting lists of subreddits that either support gun legalization or are against it;
        \item \textit{Minorities Discrimination}: identified by considering groups that promote gender/racial/sexual equality and groups with more conservative attitudes;
        \item \textit{Political Sphere}: identified by covering different US political ideologies such as Republicans, Democrats, Liberals, and Populists. 
    \end{itemize}

Data collection, users' ideology inference, and network construction are properly described in the reference paper \cite{morini2021toward}, being able to identify three users' families, protrump, antitrump, and neutral classes, that we use as our categorical attribute values.
Leveraging the original temporal network\footnote{Original network data available at \url{https://github.com/virgiiim/EC_Reddit_CaseStudy}.}, here we infer the hypergraph structure by means of all the maximal cliques.
As in the reference analysis, \cite{morini2021toward}, we consider a time window of six months when analyzing system interactions' dynamics.
Average statistics for the pairwise graphs are shown in Table \ref{tab:network_statistics}.
    
\begin{table}[t] 
\caption{Reddit Data Network statistics (averaged across semesters). size of the network in terms of nodes and edges, number of users with a Pro-Trump, Anti-Trump or Neutral leaning score. \label{tab:network_statistics}}
\centering
\scriptsize
\begin{tabular}{cccccccc}
\textbf{Topic}	& \textbf{\# nodes}	& \textbf{\# edges} & \textbf{\# Pro-Trump}  & \textbf{\# Anti-Trump}  & \textbf{\# Neutral}\\
    \hline
{{\scshape Gun Control}}&4991 &15298 &3346 &1645 &--\\
{\scshape Minorities Discrimination}&5540 & 12605 & 3318 &2222 &--\\
{\scshape Political Sphere}&4509 &7079 & 1280 &2395 &834\\
\hline
\end{tabular}
\end{table}

\subsubsection{Analytical setting}
We set a four-fold framework to analyze ideological homogeneity from different network-based perspectives as in the following:

\begin{enumerate}
    \item[i.] we promote an analysis on dyadic interactions, measuring how much users are homogeneously embedded in their pairwise ego-networks;
    \item[ii.] we shift the focus from individual users to groups, and we measure the homophily of such groups represented as hyperedges;
    \item[iii.] we come back to individual users, adopting user's point of view by measuring how much a user is embedded in the hyperedges where he/she participates;
    \item[iv.] we introduce a time-aware analysis to track stability or variations in ideological homogeneity.
\end{enumerate}

\noindent As a preliminary question, we aim to explore whether different behaviors emerge among individual users (i) and groups (ii): can high-order interactions capture patterns that are invisible to dyadic interactions?
Then, we aim to understand the role of single users in the several hyperedges where they participate (iii), as a meeting point between the two previous issues: can high-order neighborhoods capture patterns that graph ego-networks cannot?
Finally, the focus on interactions' dynamics (iv) would allow us to track stable or mutable patterns as time goes by.

\noindent It should be noted that computations in (i) and (iii) are different from (ii).
In (i) and (iii) we aim to measure the homogeneity of users' contexts with respect to the political leaning of a specific target node.
In (i), a \textit{context} is represented by the set of adjacent nodes in the ego-network of a target node, while in (iii) the context is the set of hyperedges where the target node participates in. 
We use a measure of homogeneity to estimate target nodes' similarity within nodes' own contexts.
We can use Eq. \ref{eq:homogeneity} for both pairwise and high-order nodes' ego-network, since in the former case we compute the relative frequency of the attribute values among the node's first-order neighborhood, and in the latter case we use the most frequent value as the characteristic values of a hyperedge.
Conversely, in (ii) the focus is on hyperedges' homogeneities.
Thus, we use Eq. \ref{eq:purity}, which computes the relative frequency of the most frequent attribute value within the hyperedge.

\subsubsection{Pairwise ego-networks reveal both homophilic and heterophilic users' preferences.}

\begin{figure}[]
\includegraphics[width=0.9\textwidth]{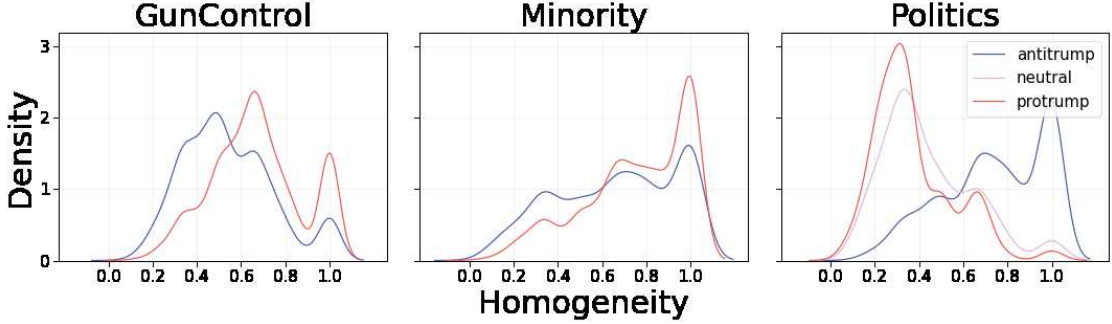}
    \caption{KDE distributions of pairwise ego-networks' homogeneity among the three different Reddit communities.}
    \label{fig:egos}
\end{figure}

Figure \ref{fig:egos} outlines graph ego-networks' homogeneities in the three topics considered. Results are aggregated over the semesters.
The analysis of pairwise interactions captures both homophilic and heterophilic patterns, telling us that such political discussions manifest heterogeneity.
For instance, in \textit{Politics}, protrump and neutral users show heterophilic behavior, while antritrump are more homogeneous.
\textit{Minority} is overall more homophilic than \textit{GunControl}, where interactions seem to be also more randomly mixed.
These observations are coherent with the analyses performed on the original data paper \cite{morini2021toward}, where in \textit{Minority} and \textit{Politics} it is more likely to observe echo-chambers -- oriented towards a protrump political leaning in \textit{Minority}, and antitrump in \textit{Politics}, while \textit{GunControl} discussions are less polarized.
\begin{figure}[]
\includegraphics[width=\textwidth]{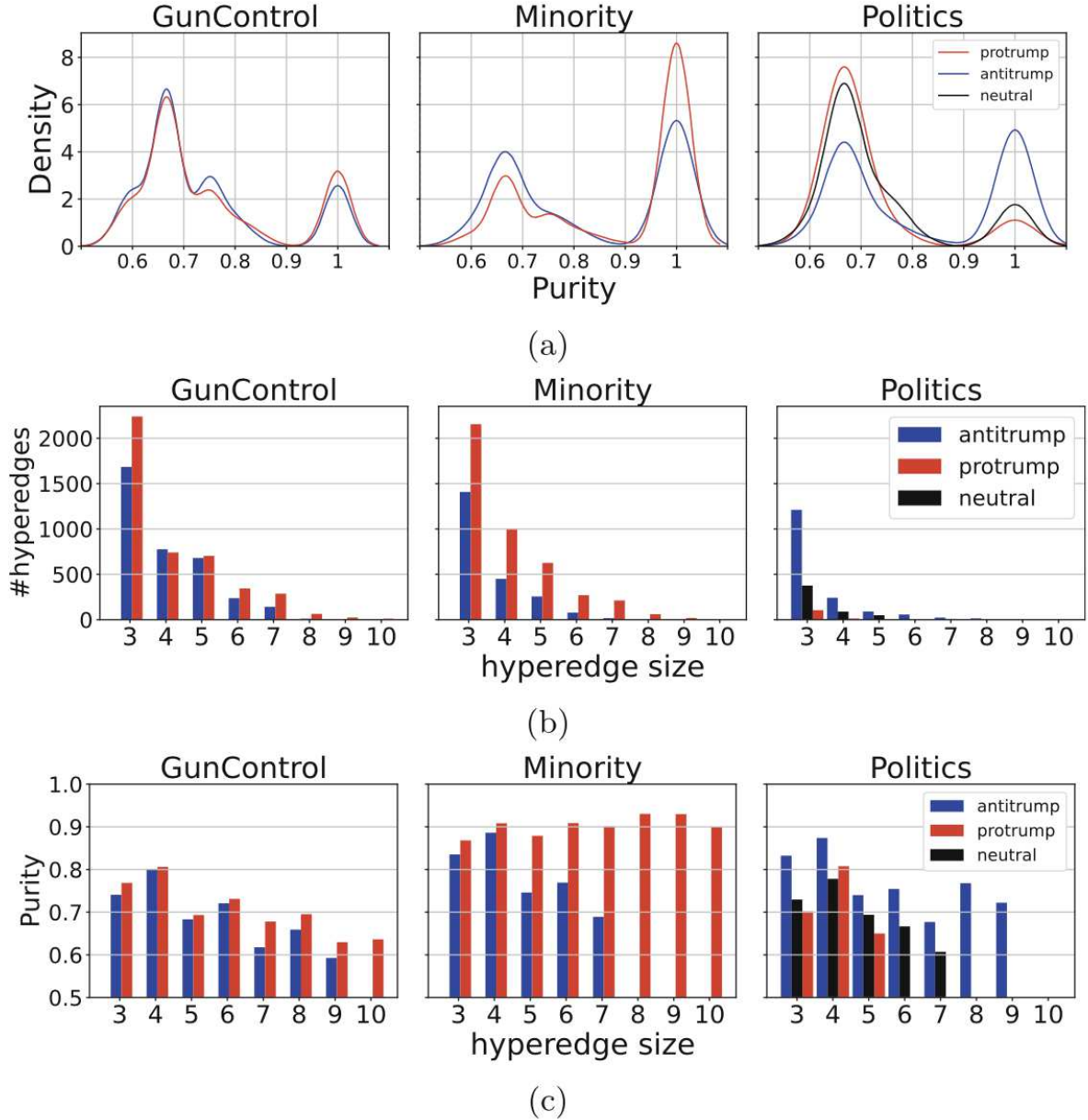}
    \caption{KDE distributions of hyperedges' purity (a), number of pure hyperedges (b), and average purity (c) in function of hyperedge size among the three different Reddit communities.}
  \label{fig:hyperhe}
\end{figure}

\subsubsection{Hyperedges' purity emphasizes heterogeneity.} 
Figure \ref{fig:hyperhe} (a) shows ideological homogeneity within the hyperedges in the three topics considered, captured by purity. Results are aggregated over the semesters.
The discussions in \textit{Minority} are the \textit{purest} ones, a result which is coherent to what already observed at the meso-scale graph-based community level in the original paper \cite{morini2021toward}:
\textit{GunControl} does not present strongly polarized communities (i.e., echo chambers) among different semesters \cite{morini2021toward} as well as it seems that only a bunch of contexts present quite perfect purity (Figure \ref{fig:hyperhe} (a), leftmost); in \textit{Minority} on average, more than half of total users are trapped in echo chambers \cite{morini2021toward}, and hyperedge purities show a quite similar pattern as well, with a tendency of protrump users to form more homogeneous groups (Figure \ref{fig:hyperhe} (a), center); also \textit{Politics} presents high homogeneity contexts, where antitrump users are more likely to form homogeneous groups (Figure \ref{fig:hyperhe} (a), rightmost).
Moreover, we analyze these patterns with respect to the hyperedge size.
Figure \ref{fig:hyperhe} (b-c) highlight, respectively, the number (b) and the average purity (c) of pure groups in function of the group size.
For instance, in \textit{Minority} we observe that only protrump pure discussions involve groups with more than $7$ participants, and that they are quite pure, $0.9$.
The same does not happen in \textit{GunControl}, while in \textit{Politics} the biggest contexts involve antitrump users only but with a lower purity than the one of protrump users in \textit{Minority}.

\begin{figure}[]
\includegraphics[width=0.9\textwidth]{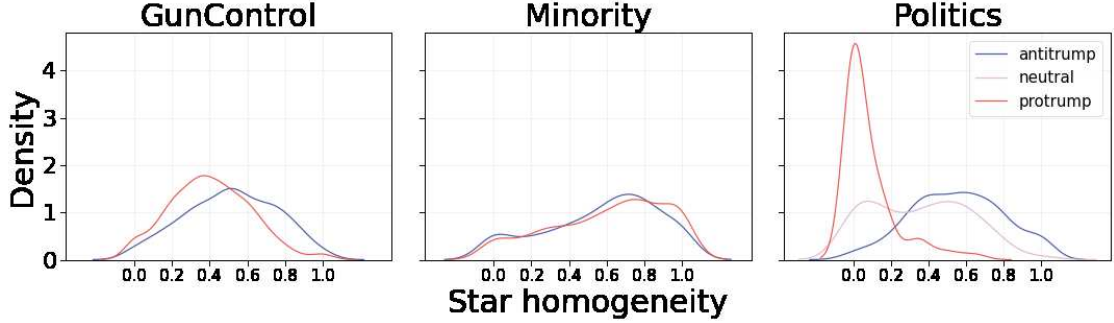}
    \caption{KDE distributions of hypergraph star ego-networks' homogeneity among the three different Reddit communities.}
    \label{fig:stars}
\end{figure}

\subsubsection{Users are involved in heterogeneous debates.}
As can be observed in Figure \ref{fig:stars}, the topics show diversified behaviors when the analysis shifts to star egos.
Indeed, there is no more trace of the heterogeneous patterns observed in Figure \ref{fig:hyperhe}(a).
The key insight, however, relates to another type of heterogeneity in user debates.
While engaging in relatively homogeneous contexts (Figure \ref{fig:hyperhe}), it seems that users find themselves in rather mixed collections of debates.
That is to say, although homophilic behavior is highlighted in most debates (i.e., hyperedges), the set of contexts a node is involved in (i.e., its star) is generally diversified with respect to ideology/political leaning.
This is especially true in \textit{GunControl}, where protrump users appear to engage in a more heterogeneous set of debates than their counterparts, as opposed to what was noted in Figure \ref{fig:hyperhe} (a).
The same holds for \textit{Politics}, which displays a peak in heterogeneous protrump stars while the antitrump ones show more homophilic behavior.
\textit{Minority}, instead, still shows strong homogeneity traits for both antitrump and protrump users, thus confirming previous observations. 

\begin{figure}[]
\includegraphics[width=0.9\textwidth]{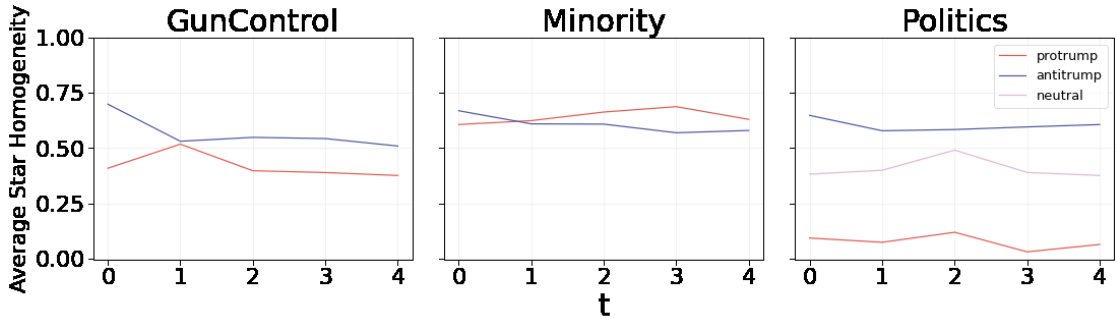}
    \caption{Average hypergraph star ego-networks' homogeneity over time among the three different Reddit communities.}
    \label{fig:startime}
\end{figure}

\subsubsection{Interactions' dynamics: users' preferences tend to be consistent in time}
\begin{table}[t] 
\caption{Reddit Data Network statistics (averaged across semesters). Nodes that stay in the network for more than two semesters/timestamps; mean and std of the consistency values for such nodes. \label{tab:consistency}}
\centering
\scriptsize
\begin{tabular}{lcccc}
\toprule
\textbf{Topic} &  \textbf{\#stay} &  \textbf{\#stay\_pct} &  \textbf{mean\_consistency (stay)} &  \textbf{std\_consistency (stay)} \\
\midrule
GunControl &   580 &       0.116 &                    0.591 &                   0.478 \\
Minority   &   735 &       0.132 &                    0.715 &                   0.441 \\
Politics   &   574 &       0.127 &                    0.748 &                   0.310 \\
\bottomrule
\end{tabular}
\end{table}

As far as the temporal dimension is concerned, a certain degree of consistency w.r.t. debates homogeneity/heterogeneity can be observed.
As a matter of fact, the average star homogeneity outlines almost-flat trends (Figure \ref{fig:startime}), indicating minor variations.
Here, \textit{GunControl} and \textit{Minority} reveal near-constant heterogeneity/homogeneity for both political alignments; lastly, \textit{Politics} displays only a small bump during the third semester concerning protrump and neutral users.
What makes this result so interesting is the fact that only $\sim 11\%$ of the nodes stay in the network for more than a semester (see Table \ref{tab:consistency}), but still coherence is observed regardless of the continuous turnover of nodes.
Such coherence is also confirmed by the \textit{Consistency} values of the remaining nodes (Table \ref{tab:consistency}, fourth column), which hint at a resilience to opinion change.

\section{Discussion and Conclusion}
\label{sec:disc}

In this work, we proposed an Attributed Stream Hypergraph (ASH) representation for taking into account both high-order relations \cite{battiston2020networks} and feature-rich information \cite{interdonato2019feature} through which to describe complex social systems.
With ASH, social phenomena represented by means of high-order interactions can also be studied together with additional information that goes beyond the network structure, namely nodes' semantics and time.
We have shown how this paradigm can be used to analyze social interactions along the (i) structural, (ii) attributive, and (iii) temporal dimensions.
The high-order architecture inherited by hypergraphs (i) allows to more realistically model social interactions which naturally occur in groups of varying sizes.
Node metadata can be used to construct node profiles (ii), which can be used to assess differences and similarities in the behaviors of different classes.
The temporal dimension (iii) can shed light on recurring patterns over time.

The novelty of ASH stands in the possibility to combine all these aspects together.
The time-respecting s-walks introduced in Section \ref{sec:ash}, for instance, open the way to a broad characterization of high-order time-changing networks, e.g., by identifying stable, densely-connected sub-hypergraphs, or to generalize dynamic centrality scores to hypergraphs \cite{simard2021computing}.
While applying the ASH framework on US-politics-bound communities on Reddit we observed strong homophilic behaviors among groups/hyperedges with respect to users’ political leaning. 
However, while focusing on the preferences of single nodes, namely on how much a target node is homogeneously embedded with respect to the representative political leaning of the groups/hyperedges it belongs to, we mostly observe a relevant decrease in nodes’ homophilic behaviors.
As a consequence, we observe that users prefer participating in contexts whose representative leaning is different than the target node’s own label, although hyperedges are strongly homophilic per se. 
Interestingly, this pattern can not be observed when looking at the pairwise ego-networks only. 

In future works, we plan to focus on the constraints that stream hypergraphs could eventually raise, such as the issues of under/overfitting social data or the robustness of the measures to missing data. 
Our findings can highlight how different temporal aggregations and graph vs. hypergraph representations deeply affect the output of analytical pipelines. 
Thus, some of the most interesting challenges in the future will be understanding the impact of different representations (e.g., graphs vs. hypergraphs), of high-order structure inference methods (e.g., via cliques~\cite{cencetti2021temporal}, overlapping communities, or other statistical methods~\cite{contisciani2022inference}), and of different measures to study mixing behaviors.
We also plan to introduce synthetic ASH generators to be used in the validation of analysis results. 

Lastly, we plan to update and maintain the \texttt{ASH} Python library, hoping it will simplify and make more accessible to researchers and practitioners feature-rich hypernetwork analysis.

\section*{Availability of data and materials}
The datasets of face-to-face interactions are available on the Sociopatterns website. \url{www.sociopatterns.org}. The Reddit data is available on GitHub \url{https://github.com/virgiiim/EC_Reddit_CaseStudy}.

\section*{Competing interests}
  The authors declare that they have no competing interests.

\section*{Author's contributions}
AF, SC, and GR designed research, performed research, and wrote the paper. 
AF and SC contributed to the acquisition, analysis, and interpretation of the experiments.
AF and GR contributed to the design of the software package rationale and its implementation. 
GR coordinated and supervised all of the research.
All authors read and approved the final manuscript.

\section*{Acknowledgements}
  This work is supported by the European Union – Horizon 2020 Program under the scheme "INFRAIA-01-2018-2019 – Integrating Activities for Advanced Communities", Grant Agreement n.871042, "SoBigData++: European Integrated Infrastructure for Social Mining and Big Data Analytics" (\url{http://www.sobigdata.eu}).

\printbibliography

@article{holme2012temporal,
  title={Temporal networks},
  author={Holme, Petter and Saram{\"a}ki, Jari},
  journal={Physics reports},
  volume={519},
  number={3},
  pages={97--125},
  year={2012},
  publisher={Elsevier}
}

@inproceedings{sarker2023generalizing,
  title={Generalizing homophily to simplicial complexes},
  author={Sarker, Arnab and Northrup, Natalie and Jadbabaie, Ali},
  booktitle={Complex Networks and Their Applications XI: Proceedings of The Eleventh International Conference on Complex Networks and their Applications: COMPLEX NETWORKS 2022—Volume 2},
  pages={311--323},
  year={2023},
  organization={Springer}
}

@article{morini2021toward,
  title={Toward a Standard Approach for Echo Chamber Detection: Reddit Case Study},
  author={Morini, Virginia and Pollacci, Laura and Rossetti, Giulio},
  journal={Applied Sciences},
  volume={11},
  number={12},
  pages={5390},
  year={2021},
  publisher={Multidisciplinary Digital Publishing Institute}
}

@article{latapy2018stream,
  title={Stream graphs and link streams for the modeling of interactions over time},
  author={Latapy, Matthieu and Viard, Tiphaine and Magnien, Cl{\'e}mence},
  journal={Social Network Analysis and Mining},
  volume={8},
  number={1},
  pages={1--29},
  year={2018},
  publisher={Springer}
}

@inproceedings{lee2021hyperedges,
  title={How do hyperedges overlap in real-world hypergraphs?-patterns, measures, and generators},
  author={Lee, Geon and Choe, Minyoung and Shin, Kijung},
  booktitle={Proceedings of the Web Conference 2021},
  pages={3396--3407},
  year={2021}
}

@article{vanhems2013estimating,
  title={Estimating potential infection transmission routes in hospital wards using wearable proximity sensors},
  author={Vanhems, Philippe and Barrat, Alain and Cattuto, Ciro and Pinton, Jean-Fran{\c{c}}ois and Khanafer, Nagham and R{\'e}gis, Corinne and Kim, Byeul-a and Comte, Brigitte and Voirin, Nicolas},
  journal={PloS one},
  volume={8},
  number={9},
  pages={e73970},
  year={2013},
  publisher={Public Library of Science San Francisco, USA}
}

@article{rossetti2018community,
  title={Community discovery in dynamic networks: a survey},
  author={Rossetti, Giulio and Cazabet, R{\'e}my},
  journal={ACM Computing Surveys (CSUR)},
  volume={51},
  number={2},
  pages={1--37},
  year={2018},
  publisher={ACM New York, NY, USA}
}

@article{divakaran2020temporal,
  title={Temporal link prediction: A survey},
  author={Divakaran, Aswathy and Mohan, Anuraj},
  journal={New Generation Computing},
  volume={38},
  number={1},
  pages={213--258},
  year={2020},
  publisher={Springer}
}

@article{cencetti2021temporal,
  title={Temporal properties of higher-order interactions in social networks},
  author={Cencetti, Giulia and Battiston, Federico and Lepri, Bruno and Karsai, M{\'a}rton},
  journal={Scientific reports},
  volume={11},
  number={1},
  pages={1--10},
  year={2021},
  publisher={Nature Publishing Group}
}

@incollection{bhagat2011node,
  title={Node classification in social networks},
  author={Bhagat, Smriti and Cormode, Graham and Muthukrishnan, S},
  booktitle={Social network data analytics},
  pages={115--148},
  year={2011},
  publisher={Springer}
}

@article{aksoy2020hypernetwork,
  title={Hypernetwork science via high-order hypergraph walks},
  author={Aksoy, Sinan G and Joslyn, Cliff and Marrero, Carlos Ortiz and Praggastis, Brenda and Purvine, Emilie},
  journal={EPJ Data Science},
  volume={9},
  number={1},
  pages={16},
  year={2020},
  publisher={Springer Berlin Heidelberg}
}

@article{musciotto2021detecting,
  title={Detecting informative higher-order interactions in statistically validated hypergraphs},
  author={Musciotto, Federico and Battiston, Federico and Mantegna, Rosario N},
  journal={Communications Physics},
  volume={4},
  number={1},
  pages={1--9},
  year={2021},
  publisher={Nature Publishing Group}
}

@inproceedings{klimt2004introducing,
  title={Introducing the Enron corpus.},
  author={Klimt, Bryan and Yang, Yiming},
  booktitle={CEAS},
  year={2004}
}

@article{zhao2018stock,
  title={Stock market as temporal network},
  author={Zhao, Longfeng and Wang, Gang-Jin and Wang, Mingang and Bao, Weiqi and Li, Wei and Stanley, H Eugene},
  journal={Physica A: Statistical Mechanics and its Applications},
  volume={506},
  pages={1104--1112},
  year={2018},
  publisher={Elsevier}
}

@article{stehle2011high,
  title={High-resolution measurements of face-to-face contact patterns in a primary school},
  author={Stehl{\'e}, Juliette and Voirin, Nicolas and Barrat, Alain and Cattuto, Ciro and Isella, Lorenzo and Pinton, Jean-Fran{\c{c}}ois and Quaggiotto, Marco and Van den Broeck, Wouter and R{\'e}gis, Corinne and Lina, Bruno and others},
  journal={PloS one},
  volume={6},
  number={8},
  pages={e23176},
  year={2011},
  publisher={Public Library of Science}
}

@article{mcpherson2001birds,
  title={Birds of a feather: Homophily in social networks},
  author={McPherson, Miller and Smith-Lovin, Lynn and Cook, James M},
  journal={Annual review of sociology},
  volume={27},
  number={1},
  pages={415--444},
  year={2001},
  publisher={Annual Reviews}
}

@inproceedings{guan2019attribute,
  title={Attribute-driven backbone discovery},
  author={Guan, Sheng and Ma, Hanchao and Wu, Yinghui},
  booktitle={Proceedings of the 25th ACM SIGKDD International Conference on Knowledge Discovery \& Data Mining},
  pages={187--195},
  year={2019}
}

@article{zhang2017when,
author = {Zhang, Fan and Zhang, Ying and Qin, Lu and Zhang, Wenjie and Lin, Xuemin},
title = {When Engagement Meets Similarity: Efficient (k,r)-Core Computation on Social Networks},
year = {2017},
issue_date = {June 2017},
publisher = {VLDB Endowment},
volume = {10},
number = {10},
issn = {2150-8097},
url = {https://doi.org/10.14778/3115404.3115406},
doi = {10.14778/3115404.3115406},
journal = {Proc. VLDB Endow.},
month = {jun},
pages = {998–1009},
numpages = {12}
}

@article{chowdhary2021simplicial,
  title={Simplicial contagion in temporal higher-order networks},
  author={Chowdhary, Sandeep and Kumar, Aanjaneya and Cencetti, Giulia and Iacopini, Iacopo and Battiston, Federico},
  journal={Journal of Physics: Complexity},
  volume={2},
  number={3},
  pages={035019},
  year={2021},
  publisher={IOP Publishing}
}

@article{pedreschi2022temporal,
  title={The temporal rich club phenomenon},
  author={Pedreschi, Nicola and Battaglia, Demian and Barrat, Alain},
  journal={Nature Physics},
  volume={18},
  number={8},
  pages={931--938},
  year={2022},
  publisher={Nature Publishing Group UK London}
}

@article{battiston2021physics,
  title={The physics of higher-order interactions in complex systems},
  author={Battiston, Federico and Amico, Enrico and Barrat, Alain and Bianconi, Ginestra and Ferraz de Arruda, Guilherme and Franceschiello, Benedetta and Iacopini, Iacopo and K{\'e}fi, Sonia and Latora, Vito and Moreno, Yamir and others},
  journal={Nature Physics},
  volume={17},
  number={10},
  pages={1093--1098},
  year={2021},
  publisher={Nature Publishing Group}
}

@article{iacopini2019simplicial,
  title={Simplicial models of social contagion},
  author={Iacopini, Iacopo and Petri, Giovanni and Barrat, Alain and Latora, Vito},
  journal={Nature communications},
  volume={10},
  number={1},
  pages={1--9},
  year={2019},
  publisher={Nature Publishing Group}
}

@article{peel2018multiscale,
  title={Multiscale mixing patterns in networks},
  author={Peel, Leto and Delvenne, Jean-Charles and Lambiotte, Renaud},
  journal={Proceedings of the National Academy of Sciences},
  volume={115},
  number={16},
  pages={4057--4062},
  year={2018},
  publisher={National Acad Sciences}
}

@article{rossetti2021conformity,
  title={Conformity: a path-aware homophily measure for node-attributed networks},
  author={Rossetti, Giulio and Citraro, Salvatore and Milli, Letizia},
  journal={IEEE Intelligent Systems},
  volume={36},
  number={1},
  pages={25--34},
  year={2021},
  publisher={IEEE}
}

@article{interdonato2019feature,
  title={Feature-rich networks: going beyond complex network topologies},
  author={Interdonato, Roberto and Atzmueller, Martin and Gaito, Sabrina and Kanawati, Rushed and Largeron, Christine and Sala, Alessandra},
  journal={Applied Network Science},
  volume={4},
  number={1},
  pages={1--13},
  year={2019},
  publisher={Springer}
}

@article{chunaev2020community,
  title={Community detection in node-attributed social networks: a survey},
  author={Chunaev, Petr},
  journal={Computer Science Review},
  volume={37},
  pages={100286},
  year={2020},
  publisher={Elsevier}
}

@article{battiston2020networks,
  title={Networks beyond pairwise interactions: structure and dynamics},
  author={Battiston, Federico and Cencetti, Giulia and Iacopini, Iacopo and Latora, Vito and Lucas, Maxime and Patania, Alice and Young, Jean-Gabriel and Petri, Giovanni},
  journal={Physics Reports},
  volume={874},
  pages={1--92},
  year={2020},
  publisher={Elsevier}
}

@article{torres2021and,
  title={The why, how, and when of representations for complex systems},
  author={Torres, Leo and Blevins, Ann S and Bassett, Danielle and Eliassi-Rad, Tina},
  journal={SIAM Review},
  volume={63},
  number={3},
  pages={435--485},
  year={2021},
  publisher={SIAM}
}

@article{chowdhary2023temporal,
  title={Temporal patterns of reciprocity in communication networks},
  author={Chowdhary, Sandeep and Andres, Elsa and Manna, Adriana and Blagojevi{\'c}, Luka and Di Gaetano, Leonardo and I{\~n}iguez, Gerardo},
  journal={EPJ Data Science},
  volume={12},
  number={1},
  pages={7},
  year={2023},
  publisher={Springer Berlin Heidelberg}
}

@article{citraro2022delta,
  title={\{$ \backslash Delta$\}-Conformity: multi-scale node assortativity in feature-rich stream graphs},
  author={Citraro, Salvatore and Milli, Letizia and Cazabet, R{\'e}my and Rossetti, Giulio},
  journal={International Journal of Data Science and Analytics},
  pages={1--12},
  year={2022},
  publisher={Springer}
}

@String{Computing = "Computing" }

@String{Computer = "{IEEE} Computer" }

@String{Springer = "Springer-Verlag" }

@article{citraro2020identifying,
  title={Identifying and exploiting homogeneous communities in labeled networks},
  author={Citraro, Salvatore and Rossetti, Giulio},
  journal={Applied Network Science},
  volume={5},
  number={1},
  pages={1--20},
  year={2020},
  publisher={SpringerOpen}
}

@article{newman2003mixing,
  title={Mixing patterns in networks},
  author={Newman, Mark EJ},
  journal={Physical review E},
  volume={67},
  number={2},
  pages={026126},
  year={2003},
  publisher={APS}
}

@inproceedings{parmentier2019introducing,
  title={Introducing multilayer stream graphs and layer centralities},
  author={Parmentier, Pimprenelle and Viard, Tiphaine and Renoust, Benjamin and Baffier, J-F},
  booktitle={International Conference on Complex Networks and Their Applications},
  pages={684--696},
  year={2019},
  organization={Springer}
}

@article{simard2021computing,
  title={Computing Betweenness Centrality in Link Streams},
  author={Simard, Fr{\'e}d{\'e}ric and Magnien, Cl{\'e}mence and Latapy, Matthieu},
  journal={arXiv preprint arXiv:2102.06543},
  year={2021}
}

@inproceedings{chiappori2021quantitative,
  title={Quantitative Evaluation of Snapshot Graphs for the Analysis of Temporal Networks},
  author={Chiappori, Alessandro and Cazabet, R{\'e}my},
  booktitle={International Conference on Complex Networks and Their Applications},
  pages={566--577},
  year={2021},
  organization={Springer}
}

@article{ribeiro2013quantifying,
  title={Quantifying the effect of temporal resolution on time-varying networks},
  author={Ribeiro, Bruno and Perra, Nicola and Baronchelli, Andrea},
  journal={Scientific reports},
  volume={3},
  number={1},
  pages={1--5},
  year={2013},
  publisher={Nature Publishing Group}
}

@inproceedings{comrie2021hypergraph,
  title={Hypergraph Ego-networks and Their Temporal Evolution},
  author={Comrie, Cazamere and Kleinberg, Jon},
  booktitle={2021 IEEE International Conference on Data Mining (ICDM)},
  pages={91--100},
  year={2021},
  organization={IEEE}
}

@article{contisciani2022inference,
  title={Inference of hyperedges and overlapping communities in hypergraphs},
  author={Contisciani, Martina and Battiston, Federico and De Bacco, Caterina},
  journal={Nature Communications},
  volume={13},
  number={1},
  pages={7229},
  year={2022},
  publisher={Nature Publishing Group UK London}
}

@article{fortunato2016community,
  title={Community detection in networks: A user guide},
  author={Fortunato, Santo and Hric, Darko},
  journal={Physics reports},
  volume={659},
  pages={1--44},
  year={2016},
  publisher={Elsevier}
}

@inproceedings{failla2023attributed,
  title={Attributed Stream-Hypernetwork Analysis: Homophilic Behaviors in Pairwise and Group Political Discussions on Reddit},
  author={Failla, Andrea and Citraro, Salvatore and Rossetti, Giulio},
  booktitle={Complex Networks and Their Applications XI: Proceedings of The Eleventh International Conference on Complex Networks and Their Applications: COMPLEX NETWORKS 2022—Volume 1},
  pages={150--161},
  year={2023},
  organization={Springer}
}

@article{gallagher2021clarified,
  title={A clarified typology of core-periphery structure in networks},
  author={Gallagher, Ryan J and Young, Jean-Gabriel and Welles, Brooke Foucault},
  journal={Science Advances},
  volume={7},
  number={12},
  pages={eabc9800},
  year={2021},
  publisher={American Association for the Advancement of Science}
}

@article{colizza2006detecting,
  title={Detecting rich-club ordering in complex networks},
  author={Colizza, Vittoria and Flammini, Alessandro and Serrano, M Angeles and Vespignani, Alessandro},
  journal={Nature physics},
  volume={2},
  number={2},
  pages={110--115},
  year={2006},
  publisher={Nature Publishing Group UK London}
}

@article{palla2007quantifying,
  title={Quantifying social group evolution},
  author={Palla, Gergely and Barab{\'a}si, Albert-L{\'a}szl{\'o} and Vicsek, Tam{\'a}s},
  journal={Nature},
  volume={446},
  number={7136},
  pages={664--667},
  year={2007},
  publisher={Nature Publishing Group UK London}
}

@article{chodrow2020annotated,
  title={Annotated hypergraphs: models and applications},
  author={Chodrow, Philip and Mellor, Andrew},
  journal={Applied network science},
  volume={5},
  number={1},
  pages={1--25},
  year={2020},
  publisher={Springer}
}

@article{christianson2020architecture,
  title={Architecture and evolution of semantic networks in mathematics texts},
  author={Christianson, Nicolas H and Sizemore Blevins, Ann and Bassett, Danielle S},
  journal={Proceedings of the Royal Society A},
  volume={476},
  number={2239},
  pages={20190741},
  year={2020},
  publisher={The Royal Society Publishing}
}

@article{ju2020network,
  title={The network structure of scientific revolutions},
  author={Ju, Harang and Zhou, Dale and Blevins, Ann S and Lydon-Staley, David M and Kaplan, Judith and Tuma, Julio R and Bassett, Danielle S},
  journal={arXiv preprint arXiv:2010.08381},
  year={2020}
}

@inproceedings{joslyn2020hypernetwork,
  title={Hypernetwork science: from multidimensional networks to computational topology},
  author={Joslyn, Cliff A and Aksoy, Sinan G and Callahan, Tiffany J and Hunter, Lawrence E and Jefferson, Brett and Praggastis, Brenda and Purvine, Emilie and Tripodi, Ignacio J},
  booktitle={International Conference on Complex Systems},
  pages={377--392},
  year={2020},
  organization={Springer}
}

@article{veldt2023combinatorial,
  title={Combinatorial characterizations and impossibilities for higher-order homophily},
  author={Veldt, Nate and Benson, Austin R and Kleinberg, Jon},
  journal={Science Advances},
  volume={9},
  number={1},
  pages={eabq3200},
  year={2023},
  publisher={American Association for the Advancement of Science}
}

\appendix 

\section{\texttt{ASH}: a Library for Attributed Stream Hypergraphs}
\label{sec:implementation}

Despite the ever-growing interest in the analysis of high-order topologies, very few software packages are available to work with hypernetworked data.
Nonetheless, most of them entirely discard temporal information -- restricting analyses to the static scenario -- and lack analytical tools to study attributive dynamics.
To address these issues, we developed \texttt{ASH}, a Python library allowing to easily handle multiadic data while retaining temporal and attributive information.
In this Section, we introduce the library rationale and describe some of the main features.\\
\textit{\textbf{Classes}}
The library's core lies in a homonymous class that offers basic functionalities related to hypergraph building, statistics, temporal information, and hypergraph transformations.
Nodes and hyperedges are assigned a unique identifier at creation (integers for nodes, strings for hyperedges), as well as initial and final temporal ids (both integers) which identify the presence of a node/hyperedge between those points in time. 
These allow to successively retrieve information about nodes and edges through time in an efficient way.
\\
Node metadata is appropriately enclosed in a separate class that represents \textit{node profiles}. 
Indeed, since nodes can have multiple attributes (i.e., inherent features) as well as other statistics (e.g., centrality scores), it is useful to enclose them in a different structure, also to handle updates and comparisons. \\
\textit{\textbf{(Dynamic) Hypergraph measures and trasformations}}
The library offers a variety of methods to compute basic node and hyperedge statistics. 
Node neighborhoods, hyperedge distributions and such can be computed both on the flattened (i.e., static, aggregated) hypergraph, or by only including hyperedges active during a specific time period.
Dynamic network measures were generalized to hypergraphs, such as node and hyperedge contribution and uniformity~\cite{latapy2018stream}.
The package provides several hypergraph transformations such as bipartite projection, dual hypergraph, and s-line graph, as well as hypergraph decomposition to graphs via clique expansion.
ASHs can also be sliced both structurally (i.e., induced sub-hypergraph) and/or temporally (i.e., hypergraph temporal slice).\\
\textit{\textbf{Paths, distances, and centralities}}.
ASH provides full support for the s-analysis framework \cite{aksoy2020hypernetwork}, which is used to generalize classic graph measures to hypergraphs.
This allows to compute paths, distances, connected components, clustering, as well as several centrality measures, and extend them along the temporal dimension. 
Time-respecting s-walks can be measured both in terms of length, weight, and duration.\\
\textit{\textbf{Attribute Analysis}}
We provide several measures to quantify mixing behaviors on dynamic hypergraphs, including but not limited to those introduced in this work. 
These quantities can characterize both nodes (e.g., measures that take into account a specific node and its surroundings) and hyperedges (e.g., measures quantifying homophilic behaviors in high interactions).\\
\textit{\textbf{Visual Analytics}}
The ASH library includes a dedicated module to facilitate the visualization of measures, node degree distribution, hyperedge size distribution, as well as time series representing the structural and attributive characteristics of the ASH through time.\\
\textit{\textbf{I/O}}
Finally, input/output facilities are handled by a dedicated module. 
In detail, it is possible to read/write node profiles from/to .csv and .json files, read/write interactions from/to csv, and also read/write the whole \texttt{ASH} from/to .json files.
\\ \ \\
All in all, the library is efficient and scales well to large hypergraphs, especially considering the complexity of the underlying model and the layers of information it operates on (nodes, relations, temporal information, and node profiles).
Such performance is primarily due to its mapping systems, namely a node-to-edge and node-to-star mappings, providing for $\mathcal{O}(1)$ access to hyperedge and node stars respectively, as well as an \texttt{ash}-native time-to-edge mapping that permits hyperedge lookups by temporal id.
\texttt{ASH} is built on top of two core libraries, \texttt{halp} and \texttt{DynetX}, to achieve these functionalities. 
Most of the hypergraph-related functionalities build on top of the \texttt{UndirectedHypergraph} class from the \texttt{halp} library, providing an efficient implementation and a wide range of utilities.
The temporal dimension is handled by \texttt{DynetX}, a library for dynamic network analysis that provides support for temporal networks, allowing modeling the evolution of networks over time. 
To date, \texttt{ASH} is the only comprehensive software package to efficiently handle hypergraph-structured data enriched with node attributes.
In fact, its competitors either do not scale well to large hypergraphs or primarily focus on other structures (e.g., directed hypergraphs).
Moreover, other libraries can only model static systems (i.e., there is no support for temporally-aware data), and lack measures and algorithms to study attribute-dependent wiring patterns.
Our library takes the best of both worlds by integrating the \textit{s}-analysis framework with node attribute analytics, and adding temporal support on top.

\end{document}